# THE BALANCING EFFECT IN BRAIN-MACHINE INTERACTION

FOTINI PALLIKARI

*Solid State Physics Department, Faculty of Physics, National and Kapodistrian University of Athens, Greece*
email: fpallik@phys.uoa.gr

**Abstract.** The meta-analysis of Intangible Brain-Machine Interaction (IMMI) data with random number generators (RNG's) (Bösch, 2006) is re-evaluated through the application of rigorous and recognised mathematical tools. The current analysis shows that the statistical average of the true RNG-IMMI data is not shifted from chance by direct mental intervention, thus refuting the IMMI hypothesis. A facet of this general statistical behavior of true RNG-IMMI data is the statistical balancing of scores observed in IMMI tests where binary testing conditions are adopted. The actual dynamics that had been supporting the elusive IMMI effect are shown to be related to the psychology of experimenters. The implications of the refutation of the IMMI hypothesis especially on associated phenomena are also discussed.

## INTRODUCTION

Twenty five years ago I started investigating the controversial world of the Intangible Brain-Machine Interaction (IMMI). It was during my sabbatical from the University of Athens, under the last of the series of Perrott-Warrick scholarships that brought me to Cambridge, UK. The scholarship supported my visit to the Mind Science Foundation (MSF) in San Antonio, Texas in 1991 where I was introduced to Dr Helmut Schmidt and his electronic random number generators (Schmidt, 1976 & 1987). These were interesting looking black boxes with buttons and switches featuring a linear or circular array of LED lights flashing randomly according to several modes of operation, figure 1. On one of its modes of operation, the linear array of lights of my IMMI device was set to flash either in one or the opposite direction. Our task was to mentally influence the progression of flashing lights in the specific direction of our choice, simply by thinking and wishing for it to happen. At the end of a run the digital display placed at its front showed a number which indicated the degree of success of our efforts. Dr Schmidt lent me one of his electronic IMMI devices so that I could assist his data collection as part of his IMMI experiment with pre-recorded targets (Schmidt et al, 1993) and also that I could additionally familiarize with it. So, I began testing the IMMI hypothesis.

The IMMI hypothesis states that "*the statistical average of random numbers is shifted away from the theoretically expected value by mental effort alone*". The difference between the observed statistical average from many IMMI tests and the expected statistical average is called the "mean-shift". According to the IMMI hypothesis the value of this mean-shift should be found significantly above zero. In other words, the hypothesis predicts that just by applying our intention, our wish or desire, we can introduce some modulation in the statistical outcomes of a random process, without the mediation of a mind-machine interface device. Such mind-machine interface devices exist to date and are used for instance to drive the mechanical parts of neuroptosthetics (Medina, 2012 & Serruya, 2002).

The present work will review the evidence in favour and against the IMMI hypothesis with true random number generators (RNG's) involving the author's own experimental observations and data treatment during the last 25 years. Yet, the major part of the analysis will involve the totality of data published during the last 35 years by independent experimenters and presented in the (Bösch et al, 2006) meta-analysis.

## THE BALANCING EFFECT IN IMMI TESTS

The adopted protocol of testing the IMMI hypothesis with the Schmidt RNG machine, consisted of one "run" of ten trials, while attempting to mentally influence the random progression of LED lights followed by another ten trials, under the same mode of operation, while allowing the machine to operate without consciously trying to mentally affect its statistics. The former type of data was named "intention" and the latter were the "no intention" data.

At a certain point there were close to 1000 trials collected, that is, about 1000 random numbers presented as z-scores in each of the two testing conditions. That made a relatively small size of a study. The z-score evaluates by how many standard errors is the average of collected numbers shifted away from the chance score. Yet, the plot of cumulative z-scores against the number of data revealed an additional intriguing element: The cumulative scores of the "intention" condition were in the direction of intention, balancing about zero mean-shift the cumulative scores of the no-intention condition. The scores were even reaching a statistically significant level in both directions, figure 2a. The data shown on the top graph of figure 2a are the intention data exhibiting positive values of the cumulative z-scores (i.e. in the direction of intention) exactly as the IMMI hypothesis predicts. The bottom data in figure 2a are the no-intention data, surprisingly keeping negative values of the cumulative z-scores (i.e. against the direction of intention) balancing the scores of the intention condition. Around about 200-250 of collected data, the cumulative z-scores appear to reach statistically significant levels: i.e. +2,5 and -3,0. The experience during these tests was imposing the impression that the mind could command the progression of lights.

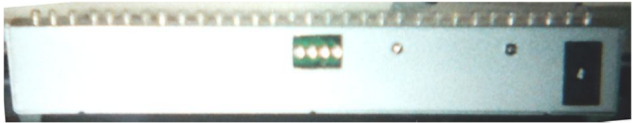

(a)

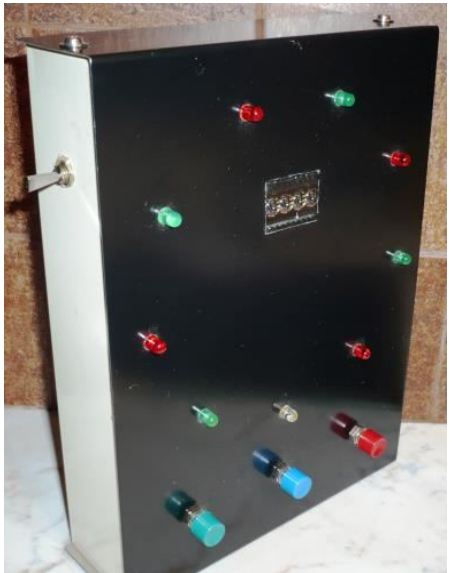

(b)

Fig. 1. Two electronic IMMI-RNG machines manufactured by Dr H. Schmidt. (a) The borrowed RNG with which the first IMMI tests were performed, through which the Balancing Effect was first observed. Its orthogonal parallelepiped shape features 31 LED lamps on its top and a digital display of the scores in the front. (b) This model of RNG device features 9 LED lamps placed circularly in front and a digital display. It was purchased by the author in 1995.

The end result of such balancing of scores was, of course, that the overall statistical average of numbers collected both under intention and no-intention conditions was not shifted away from chance, indicated in figure 2a-d by the dotted line at z=0. This bizarre statistical behavior of the random numbers generated by the Schmidt electronic RNG-IMMI machine was termed "*the balancing effect*" and presented at the 36th PA convention in Toronto, Canada in 1993 (Pallikari-Viras, 1998).

Further testing of the IMMI hypothesis (the first RNG machine had meanwhile returned back to Dr Schmidt, its chips were replaced by new ones and it was then returned to me for the follow-up studies) failed to observe the same statistical balancing of scores in a number of tests (Pallikari-Viras, 1997), figures 2b to 2d. The size of these tests was much smaller, about 250 to 500 of collected numbers in each one of them. In one of these tests, the cumulative z-scores of both conditions were observed to remain positive and not even exceeding chance levels, figure 2b. In another IMMI test a cross-over was observed around about 100 collected data. At that point of cross-over, the intention data from positive assumed negative values while the opposite was observed for the no-intention data, figure 2c. Again here the overall mean shift was not statistically different from zero. Finally, in a fourth test a similar cross-over as in the case of figure 2c was observed around about 50 collected data in each condition leaving the overall cumulative z-score again not statistically different from zero, figure 2d.

Prior to the observation of the *statistical balancing effect*, observations of similar result were independently reported. One of such cases was the "decline effect" (Rhine, 1969; Rhine 1971). The phenomenon appeared as a tendency to score below chance around the middle of a long run to render the overall score to chance and erase the previous scoring-above-chance success. Another similar report referred to the "differential effect". According to that in IMMI experiments when people are asked to produce an effect under two different conditions, they tend to score above chance in the one of them and below chance in the other (Rao, 1965). The net score would thus be a zero mean-shift. In addition, a statistical balancing of scores was also observed within the large database of Princeton Engineering Anomalies Research (PEAR) (Jahn et al, 1987). On page 119 of the book "Margins of Reality-The Role of Consciousness in the Physical World" it reads: "*When all intention data were merged with no-intention data, they yielded the theoretically expected Gaussian curve, i.e. no mean-shift.*" There were also at least, to the best of our knowledge, two reports of the *Balancing Effect* published by (Bierman et al, 1994 & Houtkooper, 2002) not prior, but this time following its first publication in related research literature.

All those independently reported cases were describing the same behavior of statistical balancing observed in various related databases, that is: *When a large enough number of random data is collected under psi or non-psi oriented conditions as part of a IMMI experiment, the overall score of the database will exhibit a zero mean-shift from chance.*

The first observation of the *Balancing Effect*, figure 2a, was simply a facet of the same fundamental law underlying all these diverse observations, i.e. the law of large numbers, that can manifest itself in more than one way in very large studies as well as in smaller datasets. According to the law of large numbers the average of the outcomes obtained from a large number of IMMI trials with RNG's gets closer to the expected value, i.e. close to chance, as more trials are collected.

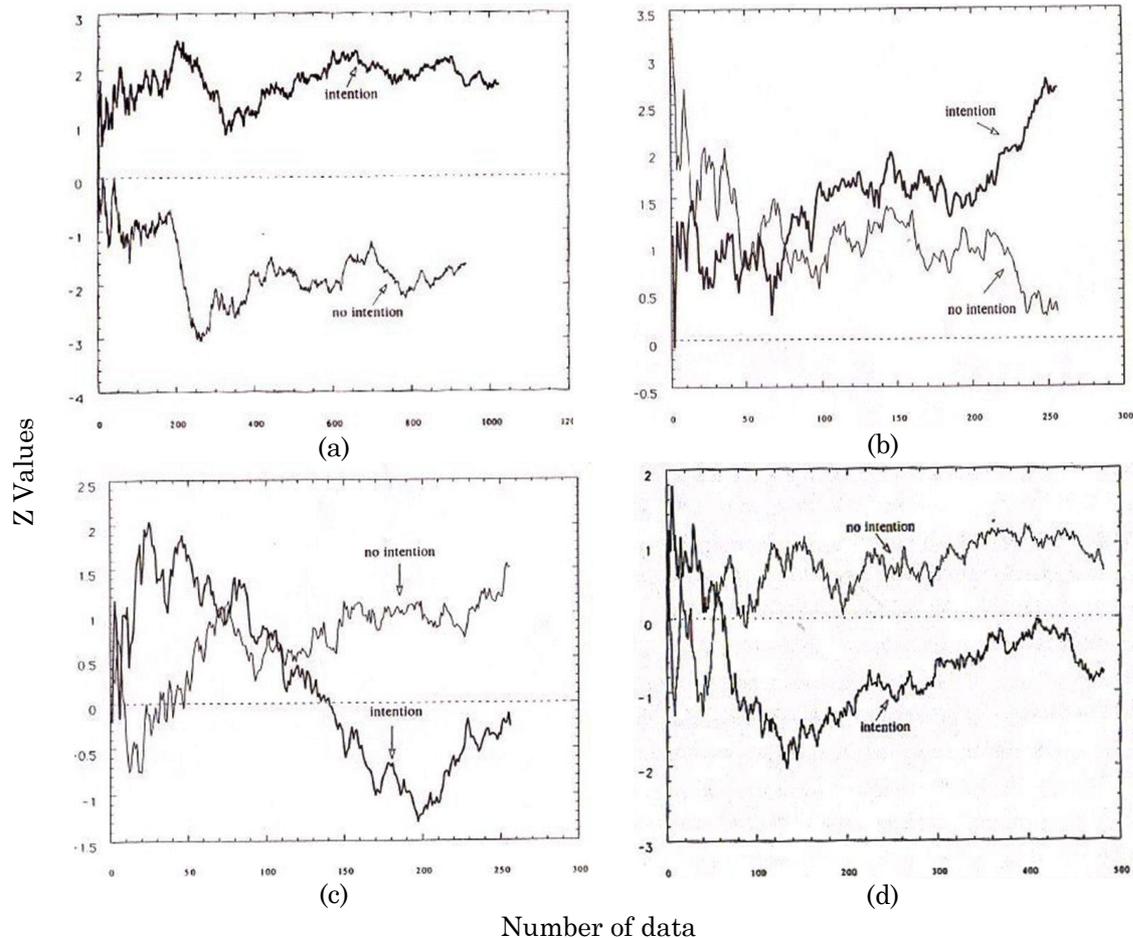

Fig. 2. Cumulative z-scores vs the number of data, N, in four MicroPK tests with an electronic Schmidt RNG (Pallikari-Viras, 1997 & 1998). (a) Scores exhibit the *balancing effect*: Intention data are positive reaching statistically significant levels and no-intention data are negative also reaching statistically significant levels; N is close to 1,000. (b) to (d): The data do not exhibit the previous statistical balancing, yet they always maintain an overall zero mean-shift. The size N in (b) and (c) is about 250 and about 500 in (d).

To support the statistical balancing observation in IMMI tests with true RNG's came a sequence of events. First, a three-leg consortium of research groups at Freiburg, Giessen, and Princeton (IMMI PortREG) was formed in 1996 to replicate prior evidence indicating that the mind could affect the random statistical behavior of electronic random number generators (Jahn et al, 1987). This large experiment failed, however, to replicate the IMMI hypothesis; the overall statistical average was found to be within chance in each one of the three experimental groups of this consortium (Jahn et al, 2000). Finally, a meta-analysis of all true RNG-IMMI data was performed triggered by this large replication failure (Bösch, 2006). Can observations that disprove the IMMI hypothesis apply not to just a part of IMMI tests, but to the large database of all IMMI tests that have ever been published? If the answer is "yes", then why do IMMI test scores exist which significantly deviate from chance?

To answer adequately questions like the above the existing IMMI evidence has been subjected to a variety of data analyses: a) the Rescaled Range Analysis which reveals the presence of long-range correlations in time series; b) the graphical representation of the RNG data on funnel plots, clearly revealing that the IMMI hypothesis is refuted while indicating the presence of correlations in the database; and c) the Markovian representation of IMMI true RNG bits which can adequately replicate the IMMI funnel plot, making itself a suitable candidate of the mechanism that introduces correlations in the database.

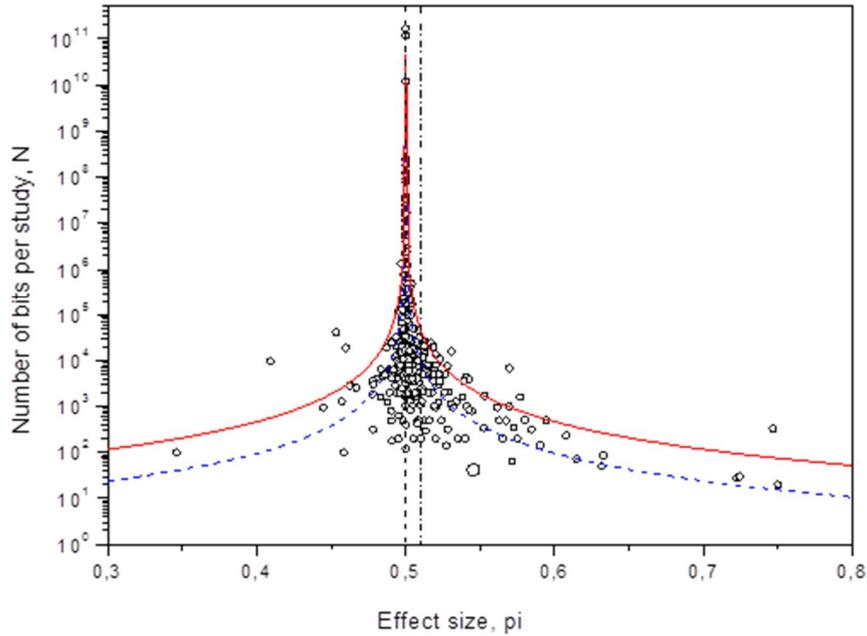

Fig. 3. Funnel plot of scores, pi, reported in IMMI data meta-analysis with true RNG's (Bösch et al, 2006). *Blue dashed curves*: The statistical border that should theoretically envelope the 95% of random plotted data. *Red solid curves*: The broadened new position of the blue dash lines that adequately envelopes the 95% of plotted data. *Dashed vertical line*: Indicates the position of the most representative effect size. *Dash-dotted vertical line*: Indicates the position of the simple statistical average of all IMMI effect sizes, $\wp = 0,51$; standard error=0,002. The standard errors of pi values are not displayed on the graph for clarity. This graph was published in (Pallikari, 2015).

## DATA ANALYSIS AND MODELLING

*The Rescale Range Analysis.*

The part of IMMI records generated at the IGPP Institute, the FAMMI (Freiburg Anomalous Mind Machine Interaction) during the three-leg IMMI consortium, was subjected to the Rescaled Range Analysis, or R/S analysis (Pallikari, 1998; Bauer, 1998). The aim of the analysis was to identify whether the IMMI records of tests arranged in time series were adequately random or correlated (Pallikari et al, 1999; Pallikari et al, 2000; Pallikari, 2001; Pallikari, 2002).

The R/S analysis estimates a parameter, the Hurst exponent, indicated by H. The records of natural phenomena are usually correlated; either by persistent correlations or by antipersistent correlations, but an overall persistence typically marks natural phenomena (Pallikari et al, 1999). The natural records may tend to persistently deviate away, either above or below chance, while the Hurst exponent ranges as: $0,5 > H \geq 1$. In the case of anti-persistence, the variation of natural records keep close to, without deviating away from, chance as compared to the random case. The Hurst exponent is found then to range as: $0 \leq H < 0,5$. In absence of correlations in a time series of natural records, the Hurst exponent is $H = 0.5$.

There were three types of FAMMI-IMMI data that were subjected to the R/S analysis; the experimental, the control and the calibration data. The "experimental data" were the records of many individual IMMI experimental sessions where operators were trying to mentally influence the behavior of the RNG device. These segments of individual tests were later merged together into the large "experimental" data time series.

|  | H | SE | N |
|---|---|---|---|
| **Experimental** | 0.521 | 0.004 | 450,000 |
| **Control** | 0.508 | 0.003 | 450,000 |
| **Calibration** | 0.505 | 0.006 | 500,000 |

Table 1. Hurst exponents, H, of IMMI data estimated by the R/S analysis. N: size of study; SE: standard error of H.

The "control data" constituted of RNG records collected usually at the end of a testing day, where the RNG was allowed to run "on its own" until the same number data was collected to equal the number of IMMI data generated on the same day. Finally, the RNG device was tested for efficiency by leaving it to run uninterrupted for very long periods of time generating the "calibration" data. The results of this R/S analysis is shown in table 1. There are persistent correlations present in the "experimental" data time series, while much weaker persistent correlations are exhibited in the "control" data. Finally, the expected randomness was exhibited by the "calibration" data, free of long-range correlations.

dependent upon the associated standard error. More precisely, IMMI experiments with very fast RNG's allow for the collection of a very large number of data. The size of experiments shown in table 1 ranges from 450 to 500 thousands of numbers. For such very large data sets the associated standard errors (se) are very low, since they are estimated as the ratio of the population standard deviation divided by the square root of the size of data. Standard error can therefore be as small as $10^{-3}$, table 1. Larger experimental sizes yield incredibly smaller standard errors and as a consequence they become the tokens of incredibly large strengths of the effect that they estimate, even if they are in absolute terms very close to null effect.

See for instance, the Hurst exponent of the control data. It appears to be close to 0.5, but it differs from it to about 2.7 standard errors and that makes the presence of persistent correlations in the control time series significant. What is the true nature of these long-range correlations, of these huge statistically significant deviations from chance (i.e. from randomness) and what is the real effect they are actually uncovering, will be

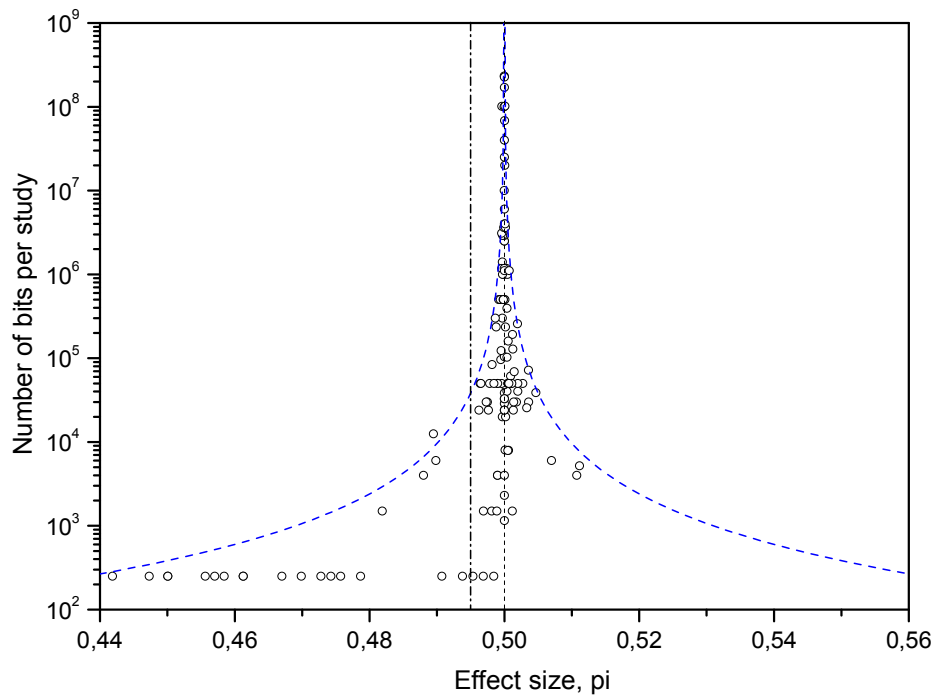

Fig. 4. Funnel plot of effect sizes obtained in control tests reported by the IMMI meta-analysis with true RNG's (Bösch, 2006). *Blue dashed curves*: 95% confidence interval as in fig.3. *Dashed vertical line*: positioned at the most representative effect size, to which the funnel plot converges and which coincides with the theoretically expected statistical average of a random binary process, 50%. *Dash-dotted vertical line*: positioned at the simple statistical average of control effect sizes $\wp = 0,495$; standard error=0,001. The standard errors of pi values are not displayed on the graph for clarity. The graph was published in (Pallikari, 2015).

At this point it becomes obvious that the evaluation of the strength of correlations in the time-series is closely elaborated in the following sections.

*The funnel plot of RNG-IMMI data.*

The following event that played an important role towards the understanding of the IMMI effect was the meta-analysis of all studies that had been performed over a period of 35 years to examine the effects of direct human intention on the concurrent output of a true RNG (Bösch et al, 2006). The authors of the meta-analysis examined the quality of the IMMI database by constructing the so-called funnel plot.

The IMMI data meta-analysis was later re-evaluated by additional analysis (Pallikari, 2015) focussing on the information revealed by the IMMI and control data funnel plots. These funnel plots are reproduced here in figures 3 and 4. Figure 3 refers to the 380 records constituting all IMMI data with true RNG's. The open circles represent the experimental data plotted as the size of study, N, against the raw proportion of hits in a study, pi. The dashed curves represent the statistical confidence interval at a 95% statistical level for random data that, as it is apparent, cannot successfully envelope the 95% of plotted data; see equations (2). The solid curves represent the new position the dashed curves should take in order to properly envelope the 95% of plotted IMMI data; see equation (3).

Provided a database is adequately large and free of biases, the scatter of data in the funnel plot is expected to be symmetric about the most representative effect size of this database, i.e. about the effect size to which the plot regresses at very large size of data, N. The shape of this graphical representation will therefore resemble that of an inverted funnel. As shown in figure 3, the most representative effect size (the actual proportion of 1's in an experimental bit sequence) in the IMMI database is 50%; i.e. chance. As it will be explained later, the totality of plotted data can be replicated by the same Markovian mathematical formulation. This feature, therefore, confirms that the totality of IMMI data with true RNG's, i.e. the small size as well as the large size studies, does not support the IMMI hypothesis, which predicts by definition a statistically significant mean-shift from chance.

A most important feature of the IMMI database is that the data do not scatter randomly about 50%, but instead their spread is broadened as compared to the spread of random data. This fact is witnessed by the inadequacy of the 95% confidence interval curves of random data to envelope the 95% of plotted IMMI records, figure 3. It is therefore concluded that the records of IMMI experiments with true RNG's are correlated.

If a large enough database is laden by the publication bias, traits of it should appear in its funnel plot. See for instance figure 3; the funnel plot of IMMI data is not symmetric. At the lower right hand side, in the region of study sizes below $N=10^5$ and effect sizes above 50%, the data scatter is more densely populated than in the left hand side. This asymmetry in shape demonstrates the presence of publication bias, which is responsible for the fact that the statistical average of all IMMI effect sizes (i. e. $\overline{p_i}_{IMMI} = 0{,}510$; standard error = 0,002), indicated by the dash dotted vertical line in figure 3, does not coincide with their most representative effect size, (i.e. 50%), indicated in figure 3 by the vertical dashed line. Publication bias in the database is due to either the reluctance of experimenters, or of journals to report the result of experiments, possibly because it contradicts the hypothesis under investigation.

A more revealing case of publication bias is shown by the funnel plot of control studies performed by the same true RNG processes that have generated the IMMI database. The control database contains only 137 data, shown in figure 4. For the case of control data, experimenters tend to withhold experimental results that they consider to challenge the effectiveness of the used RNG. Hence, the lower right hand side of the funnel plot is void of data points, where pi>50%, unlike the lower left hand side. Yet, experimental results during control tests would not have challenged the true randomness of the RNG if they had produced a score falling at the –presently void– lower right hand side region of the funnel plot under the blue dotted curve. It is expected by chance.

Here again, the statistical average of control effect sizes, $(\overline{p_i})_{control} = 0{,}495$; standard error = 0,001, marked by the vertical dash-dotted line in figure 4, does not coincide with the most representative effect size in the control database, (i.e. 50%), indicated by the vertical dashed line, as it should do. The publication bias is clearly the reason for such inconsistency.

Yet, the control data points on the funnel plot behave statistically well; they are adequately enveloped by the two 95% statistical confidence curves, the blue dashed curves in figure 4. The scores obtained during control tests in IMMI studies appear to be free of correlations.

Interestingly, the *statistical balancing* discussed previously has surfaced again in these tests, noticeable in figures 3 and 4. The statistical average of all IMMI data, $\overline{p_i}_{IMMI}$, balance the statistical average of all control data, $(\overline{p_i})_{control}$, about chance. In this case, the *balancing effect* manifests because the publication bias affects the two databases in a complimentary fashion adding up to zero mean-shift: $\overline{p_i}_{overall} = (\overline{p_i}_{IMMI} + \overline{p_i}_{control})/2 = 0.503$; standard error = 0.002. The balancing effect makes itself present here, yet again as a demonstration of the law of large numbers. Merging the two databases –IMMI and control– into a larger one is after all legitimate, since both have been generated by the same RNG processes.

In any case, the funnel plots offer a visible confirmation that the most representative effect size in each of the two databases is 50%, that is, chance. These two funnel plots, representing the entire database that has tested the effects of the mind on random physical processes by use of true RNG's, provide a concrete example of how publication bias can distort the estimated statistical average in databases creating false results.

There is another quite telling feature of the funnel plot of IMMI data shown in figure 3: The broadened scatter of data points, which had been described in a previous publication as caused by an unknown "gluing mechanism" that introduces persistence of the same RNG bit (Pallikari, 2003; Pallikari, 2004). This broadening of data scatter indicates that the records of IMMI experiments are correlated, although they represent records of independent experimental tests and as such they should have behaved statistically well as random data; i.e. they should be adequately enveloped by the blue dashed curves (the 95% confidence intervals).

*The Markovian model of RNG-IMMI data.*

The funnel plots of both IMMI and control databases clearly exhibit that the most representative effect size in both of them is null mean-shift from chance, when testing the direct influence of intention on the concurrent outcomes of a true RNG. Such an experimental result would be expected by the control database, but present in the IMMI database it provides concrete and undisputable evidence that the IMMI hypothesis is not confirmed. The mind cannot directly affect, modulate or mould physical reality.

A great puzzle is offered to interpret the presence of correlations in the IMMI records. Are such correlations the fingerprint of a different type of mind-matter interaction? In what follows it will be shown that the answer to this question is "yes", albeit not as the type of interaction that would support the IMMI hypothesis.

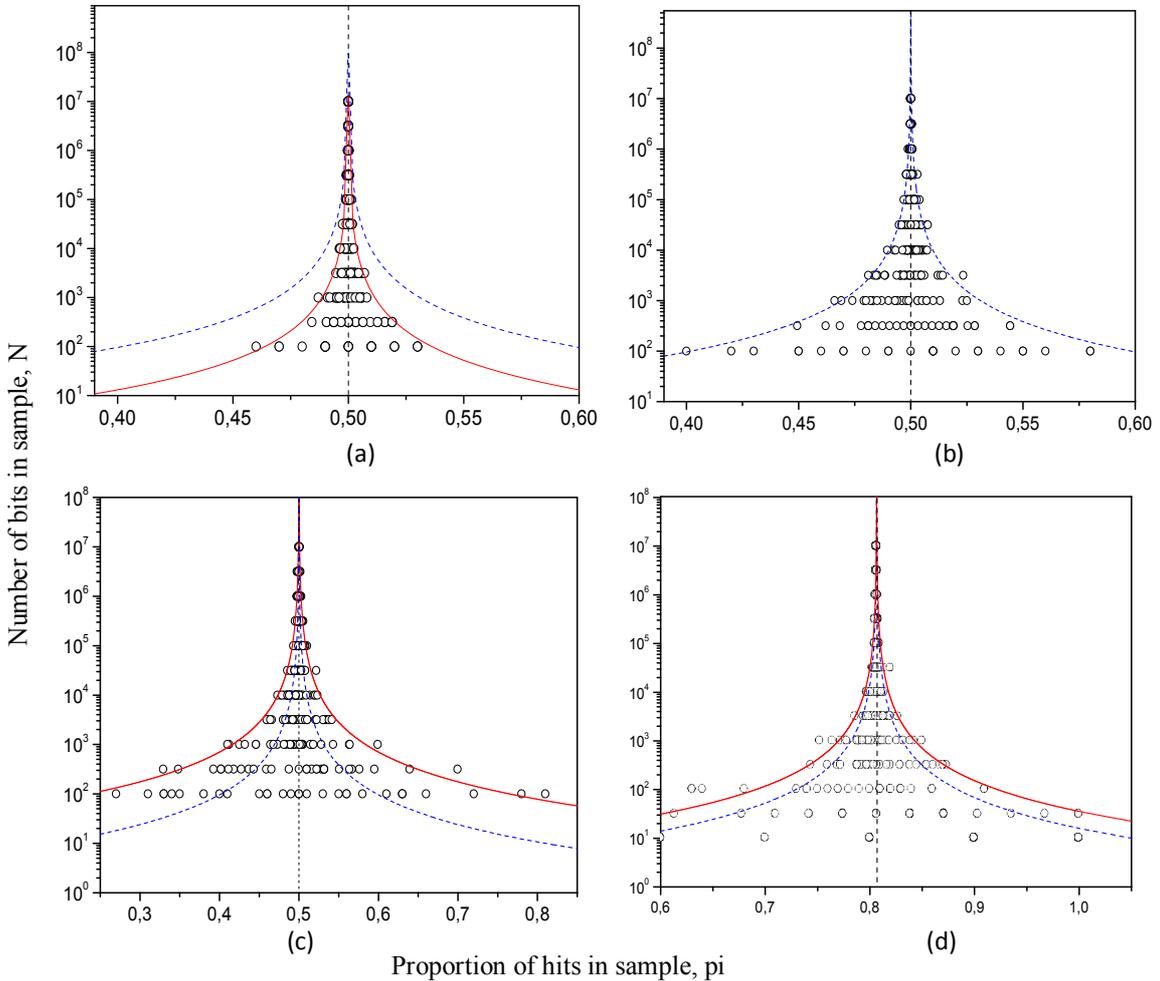

Fig. 5. Examples of funnel plots of simulated Markovian bit sequences generated at self-transition probabilities, $p_{11}$ and $p_{00}$. Open circles: The average of the proportion of 1's in 30 such samples of bits of size N. Vertical dashed line: The position $\wp$ of the most representative proportion of 1's, equation (8a); Red solid lines: 95% confidence interval of biased Markovian data, $V \neq 1$, equations (6), (7) and (8c); Blue dashed lines: 95% confidence interval of random data, V=1, equations (6), (7) and (8c). Markovian parameters: (a) $p_{11} = p_{00} = 0,12$, $\wp = 0,5$; $V = 0,37$; (b) $p_{11} = p_{00} = 0,5$; $\wp = 0,5$, $V = 1$; (c) $p_{11} = p_{00} = 0,88$, $\wp = 0,5$; $V = 2,71$; (d) $p_{11} = 0,88$ and $p_{00} = 0,5$, $\wp = 0,807$; $V = 1,49$. The graph was published in (Pallikari, 2015).

Let us first assume that an ongoing universal IMMI mechanism is underway each time a RNG bit is being registered during a IMMI trial. This is of course not only an assumption but also a generalization of a universal mechanism. It will be shown, however, that this hypothetical universal mechanism can adequately reproduce the features on the IMMI data funnel plot and as such it presents itself as a very good candidate of the actual mechanism responsible to account for the correlations in the IMMI data. More precisely, this alternative universal IMMI mechanism can adequately replicate the two characteristics of its funnel plot; the broadened scatter and the convergence to 50%, i.e. to a chance mean-shift.

This hypothetical universal IMMI mechanism works based on the following two principles: (a) *whenever a IMMI bit is generated by a true RNG (during a test somewhere in the world) the outcome depends on the previously generated bit in the same test by a universal fixed probability which is above 50%.* (b) *The above rule is independent of the type of bit; i.e. it does not distinguish between 0's and 1's.* The task will be to estimate this universal self-transition probability that replicates data present in the funnel plot of the true-RNG IMMI database.

The name given to the process that generates bits according to the above two rules is the *Markovian* process. Surely, the Markovian mechanism has refuted the IMMI hypothesis, which advocates that the mind can shift a random process in the direction indicated by intention. What the Markovian mechanism above accomplishes in the description of IMMI data is to assume a bias on the generated RNG bits equally in the intended direction as well in the opposite direction. In other words, the Markovian style of bias glues similar bits regardless of which kind they are.

Based on the above two rules an appropriate computer program has been written to simulate sequences of bits of variable length. For each sequence of length N thus generated, the average of bits, pi, has been estimated and plotted on a graph as N = f(pi). In fact, in figure 5 each data point represents the average effect size, $\overline{pi}$, of 30 sequences of size N generated by the same rules. The size of sequences, N, varies from 10 to 10.000.000 bits. The self-transition probability of bit-state "1", (i.e. the probability that a 1 will be following another 1 in the generated sequence, i.e. two 1's in a row), is $p_{11}$. Similarly, the self-transition probability of state 0, getting two 0's in a row, is $p_{00}$. The computer program can be designed to assume any relationship between $p_{11}$ and $p_{00}$, such as $p_{11} = p_{00}$ or $p_{11} \neq p_{00}$. Also, it can be arranged that $p_{11} = p_{00} > 0,5$, or $p_{11} = p_{00} < 0,5$ etc. as shown in figure 5.

As shown in figures 5a to 5c, whenever the Markovian self-transition probabilities are equal, $p_{11} = p_{00}$, the funnel plot always converges to 50%, i.e. to chance mean-shift. This replicates exactly one of the characteristic features on the funnel plot of the IMMI database. If additionally the two equal self-transition probabilities are above 50%, then the funnel plot is broadened as compared to random data, figure 5c, replicating the second characteristic feature on the funnel plot of the IMMI database.

There are other possible combinations regarding the Markovian self-transition probabilities as for instance when they are equal but lower than 50%: $p_{11} = p_{00} < 0,5$. Then the scatter of funnel plot contracts compared to that of random distribution of data, figure 5a. As a last example, figure 5d shows that the statistical average of the Markovian bits in a sequence will be shifted provided the self-transition probabilities are not equal: $p_{11} \neq p_{00}$. Then the funnel plot will converge to an effect size either above or below 50% depending on the preference of the Markovian bias for the persistence of 1's or of 0's upon bit generation, respectively. In figure 5d the preference is for 1's at a self-transition probability 88%, whereas the self-transition probability of 0's is 50%.

*Replicating the funnel plot of RNG-IMMI data.*

Mathematical formulas have so far been avoided in this discussion. The analytical mathematical description of the Markovian process can be found elsewhere (Pallikari, 2015; Papasimakis et al, 2006; Pallikari, 2008). Yet, for the sake of assisting towards a better understanding of the successful description of funnel plots by the Markovian model and also for encouraging its application on other related databases by interested researchers, some simple formulas are presented below and their application to available data is described in detail.

In equation 1(a) the letter $\wp$ denotes the most representative effect size in the database, pi; it is the effect size to which the funnel plot converges at very large sample sizes, N. Or, in other words, it is the expected proportion of hits in the IMMI studies which should coincide with the statistical average of all effect sizes in a large database free of biases. It is easily shown from equation 1a that $\wp$ is 50% when the two self-

$$\wp = \frac{1 - p_{00}}{2 - (p_{11} + p_{00})} \quad \overset{p_{11} = p_{00}}{=} \quad 0,5 \quad (a)$$

$$\sigma_0 = \sqrt{\frac{\wp(1 - \wp)}{N}} \quad \overset{p_{11} = p_{00}}{=} \quad \frac{0,5}{\sqrt{N}} \quad (b) \quad (1)$$

$$V = \sqrt{\frac{p_{11} + p_{00}}{2 - (p_{11} + p_{00})}} \quad \overset{p_{11} = p_{00} = p}{=} \quad \sqrt{\frac{p}{1 - p}} \quad (c)$$

transition probabilities are equal: $p_{11} = p_{00}$. At such condition, the Markovian process is mimicking a random process as far as the statistical average is concerned. Equation (1a) can adequately estimate the effect size $\wp$ to which a funnel plot converges for any given set of self-transition probabilities, $p_{11}$ and $p_{00}$. Simply, the self-transition probabilities of the associated Markovian process should be estimated by fitting on the funnel plot of the said database, the confidence interval curves; equations (2) and (3). The successful final result of computer simulation of Markovian data is shown in figure 5.

The sample standard deviation of effect sizes (i.e. the st. dev. of the averages of Markovian units in bits sequences of size N) whose frequency of 1's is $\wp$ and frequency of 0's is $1-\wp$, is symbolized by $\sigma_0$, in the left-hand side of equation 1b. When the two self-transition probabilities are equal, i.e. when the Markovian process is mimicking a random process, then the standard deviation acquires the familiar form as in a random process, right-hand side of equation 1b.

For as long as the generation of bits by the Markovian source fulfils the condition $p_{11} = p_{00} = 0,5$, the confidence interval of effect sizes, is given by the left-hand side of equation (2). More specifically, at the 95% statistical inference level, (z-parameter, $z_0 = 1,96$), the confidence interval takes the form of the right-hand side of equation (2)

$$pi = 0,5 \pm z_0 \cdot \frac{0,5}{\sqrt{N}} \stackrel{z_0 = 1,96}{=} 0,5 \pm \frac{0,98}{\sqrt{N}} \qquad (2)$$

The blue-dashed curves plotted on the funnel plots of experimental data, in figures 3 and 4, and on the funnel plots of simulated data, in figure 5, are based on the right-hand side of equation (2). The corresponding ones in figures 3, 5a, 5c and 5d, however, clearly do not envelope the 95% of data plotted, due to Markovian correlations present in the database which either broaden or contract the data scatter.

If the self-transition probabilities are such so that the stream of bits generated by the Markovian source are correlated, $p_{11}, p_{00} \neq 0,5$, then the confidence interval curves, $pi = f(N)$, acquire the more general form of equation (3) rather than of equation (2)

$$pi = \wp \pm z_0 \cdot \sigma_0 \cdot V = \wp \pm z_0 \cdot \sqrt{\frac{\wp(1-\wp)}{N}} \cdot V \qquad (3)$$

Equation (3) reduces to equation (2) when $p_{11} = p_{00} = 0,5$, $\wp = 0,5$ and V=1.

The variance factor, V, in equation (3) estimates the degree of deviation of the scatter of data on the funnel plot from that expected by random data. The variance factor relates to the self-transition probabilities through equation (1c). When the self-transition probabilities are equal, $p_{11} = p_{00} = p$, the variance reduces to the expression in the right-hand side of equation (1c). When the funnel plot of a database is broadened then V>1; when it is contracted then V<1, whereas in both cases the data generated by this Markovian source are correlated. When there are no correlations in the database then V=1. The variance factor, V, is estimated graphically by adjusting the confidence interval curves of equation (2) and (3) on the funnel plot. The more data present on the plot the more accurate is the graphical estimation of V and therefore the more accurate is the estimation of the two self-transition probabilities.

In the case of IMMI data the fitting the two 95% confidence interval curves yielded a variance factor V = 2,21 indicating a broadened scatter of IMMI data as compared to random data, which led to the self-transition probabilities: $p_{11} = p_{00} = 0,83$.

*Correlations in the RNG-IMMI database.*

The entirety of IMMI data collected over a period of 35 years do not support the IMMI hypothesis; direct mental effort is shown not to shift the statistical average of a random physical process. Yet, the data which testify to this clear conclusion are correlated. What is the nature of these correlations?

The success of the Markovian model to adequately replicate the features of the IMMI funnel plot makes it a good candidate of the mechanism for IMMI bit generation; the same bit has a tendency to persist regardless of type; and this is bizarre. When the RNG generates an overall statistically significant excess of 1's in a IMMI test, this indicates that intention and the outcome of the random physical process are in agreement. But, the persistent generation of 0 bits against intention in many IMMI tests to highly significant statistical levels not only invalidates the IMMI hypothesis, but also poses the question what could this correlating mechanism be that this mimicking an overall random behavior.

As one of the human traits is to imitate, a tentative mechanism for the persistent Markovian correlations observed in the IMMI database could be sought within human behavior or human psychology. Surely, the results of the "rescaled range analysis" encourage such endeavor. When the IMMI experimental records were arranged according to their publication date, persistence was also observed in their time series (Pallikari, 2015). This persistence implies that a newly published experimental result would imitate the previously published one in terms of its magnitude and sign. Successes in IMMI tests would persistently follow successes and failures would persistently follow failures to replicate the investigated IMMI hypothesis, so well balanced that overall the score of the entire database will be chance (provided publication bias is absent). What conditions could lead to such behavior?

The persistent long-range correlations in IMMI time series of experimental records imply the presence of errors that occur during the collection and treatment of experimental data that are ignored as long as the result of the current study agrees with the result of the previously published study. This behavior is well known in scientific experiments by the term "experimenter expectancy effect" (Rosenthal, 2004; Bakker et al, 2011). It refers to the unintended influence of the experimenters' hypotheses or expectations on the results of their research. Naturally, an experimental result in support of the investigated hypothesis will be reported easier if a preceding publication has reported similar results. This mimicking attitude is followed even when the result of the previously published study is against the tested hypothesis.

Notwithstanding, the probability for publicizing a success after success $p_{11}$ and the probability for publicizing a failure after failure $p_{00}$ observed in the IMMI database are equal and above 50%. The bizarre IMMI mechanism underway representing its entire database is not a direct mind-matter modulation, but a direct data modulation due to human psychology and intervention that dictates the experimenter's behavior as the human trait to err.

In a large project to estimate reproducibility within psychological data, where only a percentage slightly over 30% was reproducible (Aarts et al, 2015), it was commented: "Scientific claims should not gain credence because of the status or authority of their originator but by the replicability of their supporting evidence. Even research of exemplary quality may have irreproducible empirical findings because of random or systematic error." But, lack of reproducibility is not only observed within psychology studies; it can also be found in other fields of science (Ioannidis et al, 2009).

The invoked Markovian model provides an insight of how this error-modulated IMMI mechanism is projected at the deeper level of IMMI bits. It is making a 1-bit to be generated after a previous 1-bit more likely and a 0-bit to be generated following a 0-bit equally likely than randomness allows for. Such behavior at the bit level can unfortunately not be tested, as there is no access to the information of bit generation in each and every experiment. So, it remains just a theoretical estimation that the probability for 1-1 and for 0 - 0 bit runs is on average about 83% in IMMI time-series.

*Exploring conditions that support the IMMI hypothesis*

As the Markovian model adequately replicates the funnel plot of the RNG-IMMI data, we can reverse the question and ask: what are the required Markovian parameters displayed on the funnel plot that would testify the IMMI effect as real? What should be the results of the R/S analysis on the RNG-IMMI scores' time series for that reason? The answer to these questions can be both graphical as well as numerical. Basically, what is required to confirm the IMMI hypothesis is an IMMI bias that (A) enhances the natural repetition of bits in the direction of intention alone, i.e. only the 1's, without necessarily effecting the 0 bits. (B) Alternatively, the IMMI bias could enhance the repetition of 1's and sabotage the 0 bits by restricting their natural recurrence. We can explore these two alternatives separately.

The funnel plot that demonstrates the presence of a IMMI effect according to case (A) is graphically represented in figure 5d, in which a quite strong IMMI effect is depicted. In this example the -1- bits are generated at 88% repetition rate for the same bit (instead of 50% if the process were random) while the -0- bits are generated at 50%. The scatter of data (i.e. of the averages of such bit sequences of variable length N) is moderately broadened, V=1,49 and the funnel plot is considerably shifted from chance, centred at pi=0,807.

As it was explained in (Pallikari, 2015), the R/S analysis of the time series consisted by the averages of such bit sequences (as in a typical IMMI experiment) shows that these data are not marked by long-range correlations (represented by a Hurst exponent H=0,5) as the present correlations at the bit level are very short lived and cannot survive the process of averaging. Furthermore, a similar case displaying strong bias was discussed also in (Pallikari, 2015) where the -1- as well as the -0- bits are generated at 83%. The time series of averages of such bit sequences (that an appropriate computed program had generated) subjected to the R/S analysis confirmed that $H = 0,5$, which in turn confirms the absence of long-range correlations in their time-series as expected.

Let us look now at the numerical answer, considering an IMMI effect of gentle strength which, as in case (A), biases the 1-bits only at 55% preference, i.e. it represents an IMMI bias in the direction of intention. The associated Markovian self-transition probabilities will be, $p_{11} = 0,55$ and $p_{00} = 0,50$. The funnel plot will be centred at $\wp = 53\%$ making the shift of its centre from 50% quite noticeable. The funnel plot will be also slightly broadened, displaying a variance factor $V \simeq 1,05$. The other parameters would be: $\sigma_0 = 0,499/\sqrt{N}$, and the two confidence curves that would envelope the 95% of the treated data ($z_0 = 1,96$) would be described by $pi = 0,53 \pm 1,028/\sqrt{N}$, see equations (1) and (3).

By testing case (B) numerically, we can consider the following combination of Markovian parameters: $p_{11} = 0,55$, $p_{00} = 0,45$, $\wp = 0,55$, $\sigma_0 = 0,498/\sqrt{N}$ $V = 1$, the 95% confidence interval curves would be described by: $pi = 0,55 \pm 0,975/\sqrt{N}$. In this example the IMMI effect is favouring the bits in the direction of intention in two ways. Not only by fixing the -1- state but additionally by biasing the process to avoid the -0- bits. The funnel plot in this hypothetical case is clearly shifted away from chance, where its most representative effect size is placed at $\wp = 55\%$ in agreement with the IMMI hypothesis, although the data spread will be comparable to the case of random data, V=1.

Yet, the RNG-IMMI data funnel plot of the entire database that represents true collected during a period of 35 years is not even close to the above two numerical examples of a mild IMMI effect. Their real funnel plot is centred at $\wp = 50\%$, it is substantially broadened (V = 2,21) yielding $p_{11} = p_{00} = 0,83$ while the time-series of IMMI scores display long-range correlations (H =0,70; standard error =0,05) although they should not be correlated (Pallikari, 2015).

The above two hypothetical examples representative of evidence in favour of the IMMI effect, clearly corroborate that the actual available evidence refutes the IMMI hypothesis. Such examples also expose the basic characteristics that had been purportedly supporting the evidence for the IMMI effect. It is the psychological motivation of experimenters in certain cases to unconsciously err in order to produce experimental results that resemble the previously published IMMI tests (persistent-type of correlations in time-series).

**DISCUSSION**

*IMMI and related phenomena: The implications.*

The evidence that the IMMI hypothesis can no more be substantiated bears implications on all other related phenomena with which it shares common ground. One such phenomenon is the *macro-IMMI*; the change of position or form of a large object by mere direct mental effort. Another related phenomenon is *poltergeist* associated with the movement of objects or with noisy disturbances not of ghostly, as claimed, but of physical origin. Likewise, claims that the mind of a conscious observer interacts with the quantum state of a photon in a double-slit diffraction experiment (Radin et al, 2012 & 2013) cannot be substantiated under the current evidence which contradicts the IMMI. Claims such as all the above should seek alternative descriptions within the known laws of science, since the intangible mind-matter interaction can no longer fulfil the task. Let us have a closer look at these claims.

The IMMI hypothesis posits that the mind directly affects the inherent randomness of a physical process. Suppose that the random process refers to the thermal oscillations of atoms in a solid and that the mind could indeed directly exert a "force" to modulate their randomness, (the randomness of an incredibly large number of them), then it would make sense to even speculate the likelihood of a mental agent behind either macro–IMMI, or poltergeist phenomena. But the notion of a IMMI agent shifting a random particle motion in a desired direction as it was shown here is not validated and so is the macro–IMMI speculation as a mind-matter effect.

An alternative interpretation of the IMMI effect, was proposed invoking quantum entanglement and nonlocal correlations to bind the mind with the physical system (Lucadou v. et al, 2007; Walach et al, 2014; Lucadou v. et al, 2015). According to this interpretation the state of the mental agent is nonlocally correlated with the corresponding state of the random process as entangled parts of one system. This interpretation also implies, in reverse mode, that one could detect the state of the physical system to immediately know the mental state of a conscious observer.

Notwithstanding, correlations have indeed been identified in the large IMMI database, (their funnel plot is broadened and their time-series not random). Yet, as it was thoroughly discussed here, these long-range correlations are clearly introduced by the attitude of experimenters when either treating or publishing their data, rather than by a nonlocal mind-matter interaction. It is not so much to just identify the presence of correlations in data, but most importantly to investigate their origin.

Then it is the role of consciousness in the quantum-mechanical interpretation of the IMMI phenomena. Regarding the question posed by von Neumann–Wigner (Esfeld, 1999) whether consciousness plays an important role to collapse the quantum wavefunction during measurement of a quantum state, the opinions are divided among quantum theory scientists (Schreiber, 1995; Bierman, 2003)**.** Other than theorizing, it was to support the consciousness-related interpretation of quantum mechanics that the double-slit diffraction experiments were devised (Radin et al, 2012 & 2013), as follows.

In a double-slit diffraction experiment the wavefunction, $\Psi$, of the photon before entering the left or the right slit, exists in a superposition (or entanglement) of the two slit states. In the consciousness-related scenario adopted by (Radin et al, 2012 & 2013), consciousness, in the role of a recording device, obtains information through "mental observation" of "which path" the photon goes through, thus reducing the superposition of the two states to either the left of the right state and consequently affecting the diffraction pattern which is registered by a camera.

In double-slit experiments which slit the photon goes through is typically registered by a physical recording device. Yet, according to the IMMI scenario, the photon state is directly and mentally registered by the consciousness of the observer through an unknown and unfounded kind of "brain-physical system" interaction, as it was discussed above. The mental agent operates in both roles as the recording device of the photon path as well as the observer of the imposed consequences on the diffraction pattern.

Yet, it was shown that consciousness is not necessary to collapse the photon wavefunction in the double-slit experiment (Yu et al, 2011). In this well-formulated argument, several predictions are tried out based on available experimental results that suggest falsification of such consciousness-related interferences. In this way, the collapse-by-consciousness hypothesis in quantum measurement is rendered redundant.

## CONCLUSIONS

In year 2014 a *"call for an open, informed study of all aspects of consciousness"* was published (Cardeña, 2014) including research in all IMMI-related phenomena. The article urged towards a discussion of all evidence around such phenomena avoiding conclusions *"based solely on previously held beliefs"*, since *"scientific knowledge is provisional and subject to revision"*. The "Cardeña declaration" further stated: *"The undersigned differ in the extent to which we are convinced that the case for psi phenomena has already been made, but not in our view of science as a non- dogmatic, open, critical but respectful process that requires thorough consideration of all evidence as well as skepticism toward both the assumptions we already hold and those that challenge them"*. It is with this statement in mind, which I have also undersigned, that the present work was put together.

The current review is devoted to thoroughly evaluate the evidence concerning the IMMI hypothesis in relation to the statistical *balancing effect* that was observed in RNG IMMI tests as one facet of their fundamental behavior. The end conclusion of the overall evidence is that the IMMI hypothesis, as a modulation of the statistics of a random physical process by mere direct mental interaction is refuted.

A token for the admitted divergence of opinion as stated in (Cardeña, 2014): *"meta-analyses suggest that data supportive of psi phenomena cannot reasonably be accounted for by chance or by a 'file drawer' effect"*, is the evidence presented here, figures 3 and 4, that not only contradicts previous meta-analyses but also illustrates the reasons why their results had been fallen victim to publication bias. These two funnel plots show clearly how the file-drawer effect distorts the results of the aforementioned meta-analyses to such a degree as to shift their true statistical average, (50%, indicated by the dashed vertical line), to the fallacious statistical average (dashed-dotted vertical line). No adequate corrective statistical method has been applied, to the best of our knowledge, to bring such erratic estimate back to its correct value: to the chance meanshift.

On the basis of the current discussion can one conclude that the state of one's thoughts, wishes and intentions cannot influence future events, or others? That they cannot change the state of our health and that of others? Certainly, there exist affirmations that the aforementioned effects are indeed regularly experienced. By wondering about the real strength of our wishes and intentions, we should not underestimate the effects of our thoughts (positive or negative), of our attitudes and beliefs on our physical state, as well as that of others (Thomsen et al, 2004; Leininger, 1981).

The current analysis is thorough in the sense that it refers to the entire IMMI database being subjected to a number of recognised mathematical techniques. Alongside the refutation of the IMMI hypothesis, inclusive of small or large size studies, it further reveals the very nature of the IMMI mechanism underway; the agent that is correlating the IMMI data time series, due to parameters associated with the psychology of experimenters that contributes to the introduction of errors during experiments.

Insufficient knowledge of scientific facts may lead to wrong interpretation of the investigated phenomena in absence of a concrete scientific model. Examples are many; "metal-bending" where mental effort alone is claimed to bend a lightly touched metal or to move an object, long-distance telepathic communication with dogs, or the claim of divine messages purportedly printed on the skin of certain selected individuals. Yet, other than through trickery, one could use certain metals, e.g. Gallium, (melting point is close to 30° C) to bend a metallic object by gentle touch. Moreover, it was shown that dogs have an unbelievably sensitive olfactory system, up to 100,000-fold more sensitive as compared to human standards (Pickel et al, 2004; Beidler Walker et al, 2006). That can well account for their extraordinary feats, without necessarily ruling out the reality of other interpretations. Finally, there exists a rare skin disorder known as "Dermatographic urticaria" (Jedele et al, 1991), where the skin swells when pressed with a blunt object.

In spite of the occasional deliberate fraudulent attempts for deception due to lack of sufficient information about certain scientific facts, to which ironically even the knowledgeable can fall victim, there most likely exist real effects, e.g. telepathy and clairvoyance worth of thorough investigation. Telepathy and clairvoyance are the phenomena that refer to mental images created in the human brain that are purportedly triggered either by the thought content of another brain, or by independent environmental parameters related to places and events.

However, an adequate theory of such phenomena is lacking. In the hope that these consciousness−related phenomena will soon find a complete scientific description, perhaps through advances in neuroscience involving quantum nonlocality and quantum entanglement (Josephson et al, 1991), the words of Richard Feynman come to mind: *"The work (in science) is not done for the sake of an application. It is done for the excitement of what is found out"* (Feynman, 1998). The current work is one such attempt towards finding out the true breadth of capabilities of the human brain beyond false interpretations, misconceptions, delusions and the occasional inevitable fraud.

**Acknowledgements.** The database true RNG-IMMI records discussed here, including the associated control data, was kindly provided by Dr Holger Bösch and Dr. F. Steinkamp, co-authors of the related meta-analysis published in 2006. The department of "counselling /social and cultural research" of the Institute for Border Areas of Psychology and Mental Hygiene, IGPP, Germany, has assisted part of my search into the parapsychological literature.